\documentclass[iop,twocolappendix,numberedappendix,appendixfloats]{emulateapj}

\usepackage{graphicx,threeparttablex,longtable,array} 
\usepackage{amssymb,amsmath,multirow,gensymb}
\usepackage{enumitem,multirow,rotating}
\usepackage[caption=false]{subfig} 
\usepackage{natbib}
\newcommand{\mJybeam}{\mbox{mJy beam$^{-1}$}}
\newcommand{\vol}{\mbox{cm$^{-3}$}} 
\newcommand{\cden}{\mbox{cm$^{-2}$}} 
\newcommand{\kms}{\mbox{km s$^{-1}$}}
\newcommand{\um}{\mbox{$\mu$m}}

\newcommand{\Msun}{\mbox{M$_{\odot}$}}
\newcommand{\cmg}{\mbox{cm$^2$ g$^{-1}$}}

\newcommand{\NHH}{\mbox{N(H$_2$)}}

\newcommand{\CCO}{\mbox{$^{13}$CO}}
\newcommand{\COO}{\mbox{C$^{18}$O}}
\newcommand{\DCOp}{\mbox{DCO$^{+}$}}

\newcommand{\PI}{\mbox{$\mathcal{P}_I$}}
\newcommand{\sPI}{\mbox{$\sigma_{\mathcal{P}I}$}}
\newcommand{\PF}{\mbox{$\mathcal{P}_F$}}

\begin{document}

\title{Dust Polarization Toward Embedded Protostars in Ophiuchus with ALMA. I. VLA 1623}

\author{Sarah I. Sadavoy\altaffilmark{1$\dagger$}, 
Philip C. Myers\altaffilmark{1},
Ian W. Stephens\altaffilmark{1},
John Tobin\altaffilmark{2,3},
Beno\^{i}t Commer\c{c}on\altaffilmark{4},
Thomas Henning\altaffilmark{5},
Leslie Looney\altaffilmark{6},
Woojin Kwon\altaffilmark{7,8},
Dominique Segura-Cox\altaffilmark{9},
Robert Harris\altaffilmark{6}
	 }

\footnotetext[$\dagger$]{Hubble Fellow}
	 
\altaffiltext{1}{Harvard-Smithsonian Center for Astrophysics, 60 Garden Street, Cambridge, MA, 02138, USA}
\altaffiltext{2}{Homer L. Dodge Department of Physics and Astronomy, University of Oklahoma, 440 W. Brooks Street, Norman, OK 73019, USA}
\altaffiltext{3}{Leiden Observatory, Leiden University, P.O. Box 9513, 2300-RA Leiden, The Netherlands}
\altaffiltext{4}{Universit\'{e} Lyon I, 46 All\'{e}e d'Italie, Ecole Normale Sup\'{e}rieure de Lyon, Lyon, Cedex 07, 69364, France}
\altaffiltext{5}{Max-Planck-Institut f\"{u}r Astronomie (MPIA), K\"{o}nigstuhl 17, D-69117 Heidelberg, Germany}
\altaffiltext{6}{Department of Astronomy, University of Illinois, 1002 West Green Street, Urbana, IL, 61801, USA}
\altaffiltext{7}{Korea Astronomy and Space Science Institute (KASI), 776 Daedeokdae-ro, Yuseong-gu, Daejeon 34055, Republic of Korea}
\altaffiltext{8}{Korea University of Science and Technology (UST), 217 Gajang-ro, Yuseong-gu, Daejeon 34113, Republic of Korea}
\altaffiltext{9}{Centre for Astrochemical Studies, Max-Planck-Institute for Extraterrestrial Physics, Giessenbachstrasse 1, 85748, Garching, Germany}


\date{Received ; accepted}

\begin{abstract}
We present high resolution ($\sim 30$ au) ALMA Band 6 dust polarization observations of VLA 1623.  The VLA 1623 data resolve compact $\sim 40$ au inner disks around the two protobinary sources, VLA 1623-A and VLA 1623-B, and also an extended $\sim 180$ au ring of dust around VLA 1623-A.  This dust ring was previously identified as a large disk in lower-resolution observations.  We detect highly structured dust polarization toward the inner disks and the extended ring with typical polarization fractions $\approx 1.7$\% and $\approx 2.4$\%, respectively.  The two components also show distinct polarization morphologies.  The inner disks have uniform polarization angles aligned with their minor axes.  This morphology is consistent with expectations from dust scattering.  By contrast, the extended dust ring has an azimuthal polarization morphology not previously seen in lower-resolution observations.  We find that our observations are well-fit by a static, oblate spheroid model with a flux-frozen, poloidal magnetic field.  This agreement suggests that the VLA 1623-A disk has evidence of magnetization.  Alternatively, the azimuthal polarization may be attributed to grain alignment by the anisotropic radiation field.  If the grains are radiatively aligned, then our observations indicate that large ($\sim 100$ \um) dust grains grow quickly at large angular extents.  Finally, we identify significant proper motion of VLA 1623 using our observations and those in the literature.  This result indicates that the proper motion of nearby systems must be corrected for when combining ALMA data from different epochs.
\end{abstract} 


\section{Introduction\label{Intro}}

Magnetic fields are expected to impact the formation of protostars and their disks.  Theoretical studies show that angular momentum of infalling material is transported outward by magnetic fields.  This process, called magnetic braking, can greatly suppress the formation of disks and companion stars \citep[e.g.,][]{Machida05, PriceBate07}.  A large body of work has examined ways to mitigate magnetic braking so that Keplerian disks and binaries are able to form early in the star formation process when there is a large mass reservoir.  These studies include misaligned magnetic fields relative to the rotation axis of the collapsing core \citep[e.g.,][]{HennebelleFromang08, HennebelleCiardi09, Machida11}, non-ideal magnetic hydrodynamic processes such as ambipolar diffusion and Ohmic dissipation \citep[e.g.,]{Tomida15, Masson16, Hennebelle16, Vaytet18}, and turbulence \citep[e.g.,][]{Seifried13, Joos13, Gray18}.

Dust polarization is commonly used as a tracer of magnetic fields in star-forming regions.  Non-spherical dust grains in molecular clouds are expected to spin due to radiative alignment torques (RATs) from an anisotropic radiation field.  In the presence of an external magnetic field, these spinning grains will preferentially align with their short axes parallel to the magnetic field lines, producing polarized dust extinction from background starlight that is parallel to the magnetic field or polarized thermal dust emission that is orthogonal to the field \citep[e.g.,][]{DolginovMitrofanov76, ChoLazarian07, Andersson15}.   Polarization from thermal dust emission in particular has been used to infer the plane-of-sky magnetic field structure from the scale of our Galaxy \citep[e.g.,][]{PlanckB15, Soler17} to the scales of individual star-forming cores and protostellar envelopes \citep[e.g.,][]{Matthews09, Hull14, Maury18}.

The first studies to resolve dust polarization in protostellar disks from (sub)millimeter observations \citep{Rao14, Stephens14, SeguraCox15} were initially associated with magnetic fields.  These results, however, are ambiguous results as dust polarization can arise from several alternative mechanisms that can be significant in disks.   These mechanisms include Rayleigh self-scattering from dust grains in an anisotropic radiation field  \citep[e.g.,][]{Kataoka15, Pohl16, Yang16} and grain alignment from radiative torques themselves along the flux gradient of the radiation field \citep[e.g.,][]{Tazaki17}.  Indeed, high-resolution dust polarization observations of circumstellar disks with ALMA \citep[e.g.,][]{Kataoka16, Kataoka17, Stephens17, Lee18, Cox18} show evidence of these alternative mechanisms.  These studies suggest that dust polarization detections on disk-scales may instead trace a combination of mechanisms, although we still have relatively small and inhomogeneous samples of sources.
 
 With our ALMA study of embedded stars in Ophiuchus, we obtain a larger sample of dust polarization toward embedded objects.  This study includes all embedded (Class 0 and Class I) protostars in the nearby \citep[120 pc;][]{Loinard08} Ophiuchus molecular cloud with ALMA in full dust polarization in Band 6 (1.3 mm) at a high angular resolution of $\sim 0.25\arcsec$ ($\sim 30$ au).  These observations trace dust polarization in all these systems on scales that include their disks and inner envelopes.  This study is both uniform and unbiased, and offers an unprecedented view of dust polarization toward embedded protostars.

In this paper, we present an initial study on VLA 1623, the canonical Class 0 system \citep{Andre93}.  VLA 1623 is a triple star system \citep{Looney00}, with a pair of sources, VLA 1623-A and VLA 1623-B, separated by roughly $1.17$\arcsec\ and a third component, VLA 1623-W, at a distance of $\sim 10$\arcsec\ west from the tight binary.  Previous high-resolution observations of VLA 1623 found extended dust and rotating gas emission around VLA 1623-A that has been attributed to a large Keplerian disk \citep{Murillo13}.  The system also has a well collimated outflow \citep{Andre90}.  Dust polarization from single-dish telescopes \citep{Matthews09, Dotson10} and coarse ($\sim 4$\arcsec) resolution CARMA observations \citep{Hull14} suggest an inferred magnetic field that is perpendicular to this outflow.  

With our ALMA data of VLA 1623, we detected highly structured dust polarization toward VLA 1623.  We identify two distinct polarization morphologies - a uniform pattern associated with the compact protostellar disks around each source and an azimuthal pattern associated with the extended emission around VLA 1623-A.  In Section \ref{data}, we describe our observations.  In Section \ref{results}, we show the polarization results and in Section \ref{pol_section}, we discuss the polarization morphologies.  In Section \ref{discussion}, we compare our observations qualitatively to models and to previous observations of VLA 1623.   Finally, in Section \ref{summary}, we summarize our main conclusions.


\section{Observations}\label{data}

VLA 1623 was observed in full polarization with ALMA in Band 6 (1.3 mm, 233 GHz) on 20 May, 11 July, and 13 July 2017 as part of a larger Cycle 3 (2015.1.01112.S) polarization survey.  In brief, the survey observed 26 embedded protostellar systems (Class 0 and Class I) in the Ophiuchus molecular cloud using 28 pointings in the C40-5 configuration with 36 antenna and a maximum baseline of 1.12 km (May observations) or 2.65 km (July observations).  Two sources, VLA 1623 and IRAS 16293-2422, were observed with two fields centered on each of the their wide binary components.  The protostars were selected from \citet{Enoch09}, \citet{Evans09}, and \citet{ConnelleyGreene10} which sampled young embedded stars in Ophiuchus.  We use the infrared spectral index \citep[$\alpha > -0.3$;][]{Evans09} to select only Class 0 and Class I sources.

All 28 pointings were observed during each scheduling block over at least 3 hours for sufficient parallactic angle coverage to calibrate the polarization D-terms.  The total time on each target field is 7.2 minutes.  For VLA 1623, we have two pointings; one pointing is centered on the central A and B sources and the second pointing is on the western W source \citep[e.g.,][]{Looney00}.  For this paper, we focus on the observations of the central A and B sources\footnote{We do not mosaic the two VLA 1623 pointings.  The western source is located $\sim 10$\arcsec\ from the central A and B sources, which places it outside of the inner third of the primary beam ($\sim 23$\arcsec).  Mosaics for such large separations are not recommended by the ALMA Science Center}.

The data are calibrated using J1517-2422 for bandpass, J1625-2527 for phase, and J1549+0237 for polarization leakage.  For flux calibration, we use J1517-2422 for the 20 May and 13 July observations and J1733-1304 for the 11 July observations. The data are calibrated manually using CASA 4.7.2. Full details on the entire survey will be provided in a future paper.


We image the VLA 1623 observations using \emph{clean} interactively.   Since VLA 1623 is a bright source (S/N $\gg 25$), we self calibrate our observations.  We test several combinations of phase-only self calibration and phase with amplitude self calibration.  For each test, we (1) perform self calibration two or three times using successively smaller time intervals, (2) visually check the amplitude profile of VLA 1623 after applying the solutions, and (3) re-image the data to compare signal and sensitivity from the previous round.   The phase-only self calibrations maintain a constant amplitude profile and improve the map sensitivity.  The cases with amplitude self calibration, however, introduce spikes at large uv-distances and lose source amplitudes at short uv-distances.  Thus, for our final maps, we use three rounds of phase-only self calibration.

We perform one final, deep iteration of clean and a primary beam correction on each of the four Stokes maps using our self-calibrated data.  For Stokes I, we use \emph{clean} in interactive mode to manually identify regions with bright emission.  We also use the multi-scale option to recover the extended emission better.  For the Stokes Q, U, and V observations, we run \emph{clean} non-interactively and without multi-scale.  The final map sensitivities are 0.071 \mJybeam, 0.027 \mJybeam, 0.027 \mJybeam, and 0.026 \mJybeam\ for the Stokes I, Q, U, and V maps, respectively.    The sensitivity of the Stokes I map is dynamic range limited, resulting in its higher rms value.  Our map resolution is $0.27\arcsec \times 0.21\arcsec$ or $\sim 30$ au at the distance of Ophiuchus.


\section{Results}\label{results}

\begin{figure}[h!]
\includegraphics[width=0.475\textwidth,trim=1pt 1pt 1pt 1pt,clip=true]{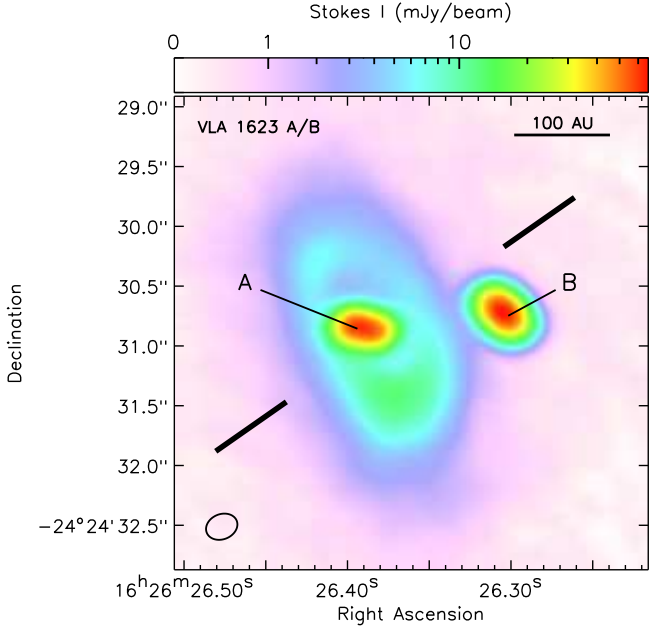}
\caption{ALMA 1.3 mm Stokes I observations of VLA 1623.  The field is centered on the VLA 1623A and VLA 1623B sources.  The thick black lines show the direction of the collimated outflow lobes \citep[position angle is -55\degree;][]{Santangelo15}.  The synthesized beam size is in the lower-left corner.  \label{stokesI}}
\end{figure}

Figure \ref{stokesI} shows the Stokes I map of VLA 1623-A and VLA 1623-B.  The phase center is $\alpha$ $=$ 16:26:26.351 and $\delta =-$24:24:30.57 (J2000).  Note that the background image follows a logarithmic color scale.  Both VLA 1623-A and VLA 1623-B are well detected and resolved in our ALMA Stokes I observations.  We also resolve extended, asymmetric emission around VLA 1623-A.  This extended was previously detected at 1.3 mm with ALMA in early Cycle 0 observations at a lower resolution of $\sim 0.65$\arcsec\ and also coincides with red- and blue-shifted \COO\ (2-1) emission that appears to trace a Keplerian disk \citep{Murillo13}.  With our high-resolution observations, the extended emission has a ring-like morphology, although we detect no dips in an azimuthally-averaged intensity profile.  We call this region an ``extended dust ring'' to distinguish it from the more compact emission centered on VLA 1623-A and VLA 1623-B.

\begin{figure*}
\centering
\includegraphics[trim=0mm 4cm 0mm 9mm,clip=true,width=0.95\textwidth]{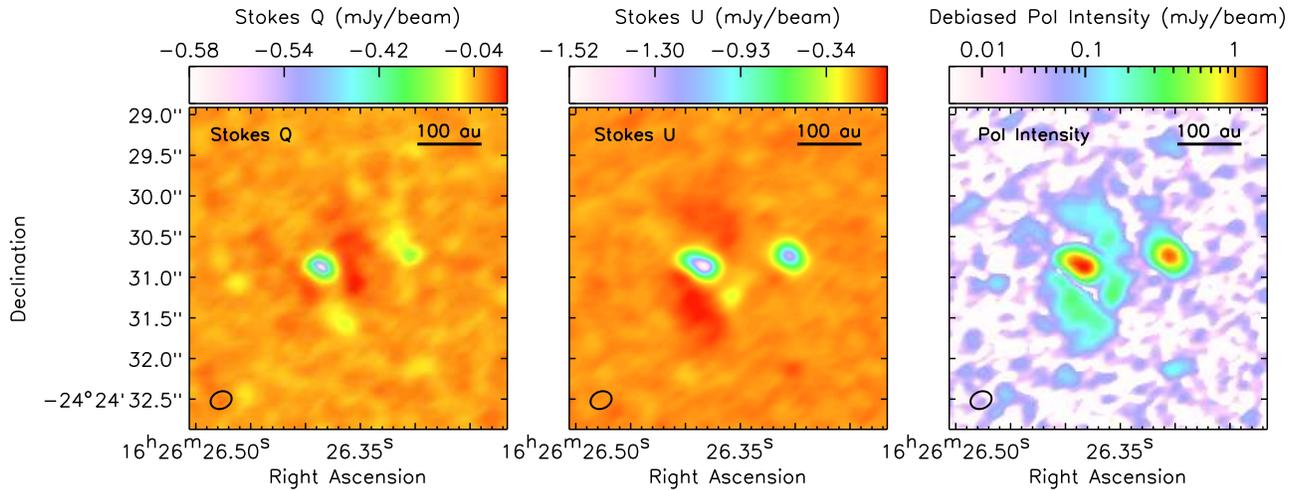}
\caption{VLA 1623 in Stokes Q (left), Stokes U (middle), and debiased polarized intensity (right).  Note that the three maps use different logarithmic scaling.    \label{stokesQU}}
\end{figure*}

Figure \ref{stokesQU} shows the Stokes Q and U maps and the polarized intensity map.  Note that the three figures use different logarithmic scales to highlight extended features.  VLA 1623-A and VLA 1623-B are each well detected in the individual Stokes Q and U maps and in polarized intensity.  The extended ring, however, is less prominent and appears partially traced in polarized intensity. The Stokes V map does not have any detections.  

The polarized intensity map in Figure \ref{stokesQU} corresponds to the \emph{debiased} polarized intensity.  In brief, polarized intensity is measured from the quadrature sum of the Stokes Q and U maps ($\PI_{,uncorr} = \sqrt{Q^2+U^2}$).   This quadrature sum always produces a positive value, even though the Stokes Q and U maps may be negative (e.g., see Figure \ref{stokesQU}).   The resulting polarized intensities are then biased to higher values as sign differences in Stokes Q and U are not translated to polarized intensity.   This bias must be removed from observations to determine the true polarized intensity.

The most common approach to debias dust polarization observations follows a basic maximum likelihood characterization \citep{SimmonsStewart85, Vaillancourt06}:

\begin{equation}
\PI = \sqrt{Q^2+U^2 - \sPI^2} \label{debiasEq},
\end{equation}
where $Q$ and $U$ are the Stokes Q and U intensities at each pixel and $\sPI$ is the noise in the polarization map.  We assume $\sPI \approx \sigma_{Q} \approx \sigma_U$ for simplicity.  Since the maximum likelihood method is reliable only for well-detected dust polarization, we limit our analysis to only those data with $\PI/\sPI >4$  \citep{Vaillancourt06}. 

We calculate the position angle of the polarization vectors, $\theta$, with
\begin{equation}
\theta = \frac{1}{2}\tan^{-1}\frac{U}{Q}, \label{ang_eq}
\end{equation}
and the polarization fraction, \PF, with
\begin{equation} 
\PF = \frac{\PI}{I}.
\end{equation}
All polarization position angles are measured from $-90\degree$ to 90\degree, North to East.  The 180$\degree$ ambiguity in the position angle from Equation \ref{ang_eq} is resolved from identifying the appropriate Cartesian quadrants based on the values of U and Q.  For $\PI/\sPI > 4$, the uncertainty in polarization position angle is $\lesssim 7$\degree\ \citep[e.g.,][]{Hull14}.  We assume a 1\ $\sigma$ polarization fraction uncertainty of 0.1\%, which is appropriate for extended emission within the inner third of the primary beam.

\begin{figure*}
\centering
\includegraphics[width=0.95\textwidth,trim=1pt 1pt 1pt 1pt,clip=true]{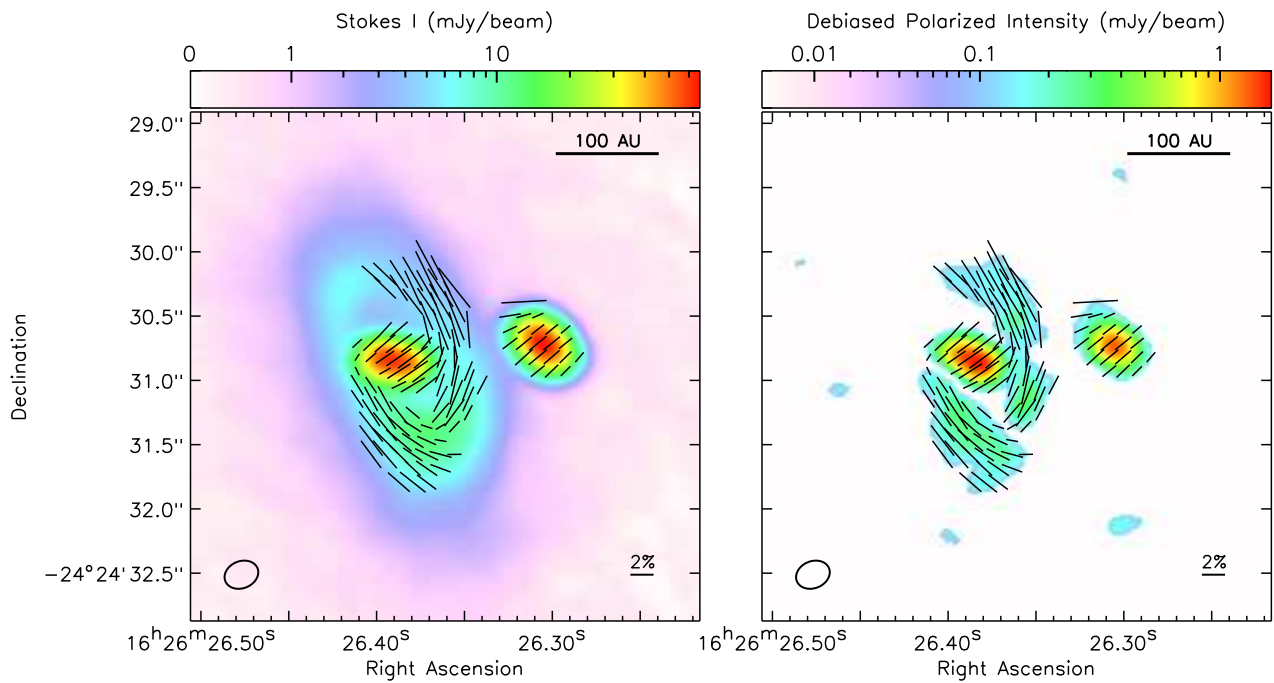}
\caption{The polarization morphology of VLA 1623.  The e-vectors correspond to only those data with $I > 3\ \sigma_I$ and $\PI > 4\ \sPI$.  The e-vectors are also scaled by the (debiased) polarization fraction.  A reference scale of $\PF = 2$\%\ is given in the lower-right corners.  The background images are (left) Stokes I as shown in Figure \ref{stokesI} and (right) polarized intensity map as shown in Figure \ref{stokesQU} for $\PI > 4\ \sPI$.     \label{polarization}}
\end{figure*}

Figure \ref{polarization} shows the observed polarization morphology of VLA 1623 overlayed on maps of Stokes I and polarized intensity.  The polarization vectors\footnote{Dust polarization gives a position angle only, whereas true vectors also have a direction.  Some studies use the term ``half-vector'' \citep[e.g.,][]{Ward-T17} or ``polar'' \citep[e.g.,][]{Hull14}.  We prefer the term ``vector'' and use it here.} are scaled by the polarization fraction with a reference of 2\%\ given in the lower-right corner.  Hereafter, we use the term ``e-vector'' to correspond to the observed polarization vectors (e.g., no rotation has been applied).

\section{Polarization Morphology} \label{pol_section}

Figures \ref{stokesQU} and \ref{polarization} shows substantial polarization structure toward VLA 1623 for the first time.  In particular, the e-vectors appear to follow different morphologies for the inner compact objects around each protostar and the extended dust ring.  The inner compact objects show relatively uniform polarization position angles, whereas the extended dust ring appears to have an azimuthal polarization pattern that traces the general curvature of the ring over roughly three-quarters of its extent.   Hereafter, we call the compact density peaks at the position of each protostar ``inner protostellar disks'' although we cannot be sure both of these structure are genuine disks.  For VLA 1632-A, there is evidence that its compact dust emission may be tracing a Keplerian disk \citep{Murillo13}.  For VLA 1623-B, however, we argue that it is also a disk candidate based on its compact shape and density constrast.

\begin{figure}[h!]
\includegraphics[width=0.475\textwidth,trim=1pt 1pt 1pt 1pt,clip=true]{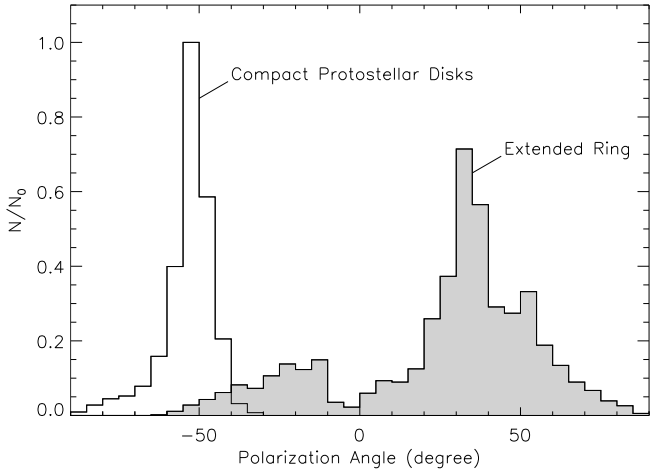}
\caption{Histograms for polarization position angle.  The solid histogram corresponds to angles associated with the extended dust ring and the open histogram corresponds to angles associated with the compact, inner protostellar disks.  For definitions on how the regions were defined see the text and Figure \ref{masks}.  \label{angle_hist}}
\end{figure}

Figure \ref{angle_hist} compares the distribution in position angle for the inner protostellar disks (open histogram) and the extended ring (filled histogram).  We describe how these two regions were differentiated below.  The two histograms have very different profiles with little overlap.  The two inner disks have a narrow distribution in angle that peaks around -50$\degree$, whereas the extended ring has a broader distribution that peaks at roughly 30$\degree$.  This figure illustrates that the inner protostellar disks and the extended ring have distinctly different polarization morphologies, which could indicate different mechanisms producing their polarization. 

\begin{figure*}
\centering
\includegraphics[trim=0mm 4cm 0mm 9mm,clip=true,width=0.95\textwidth]{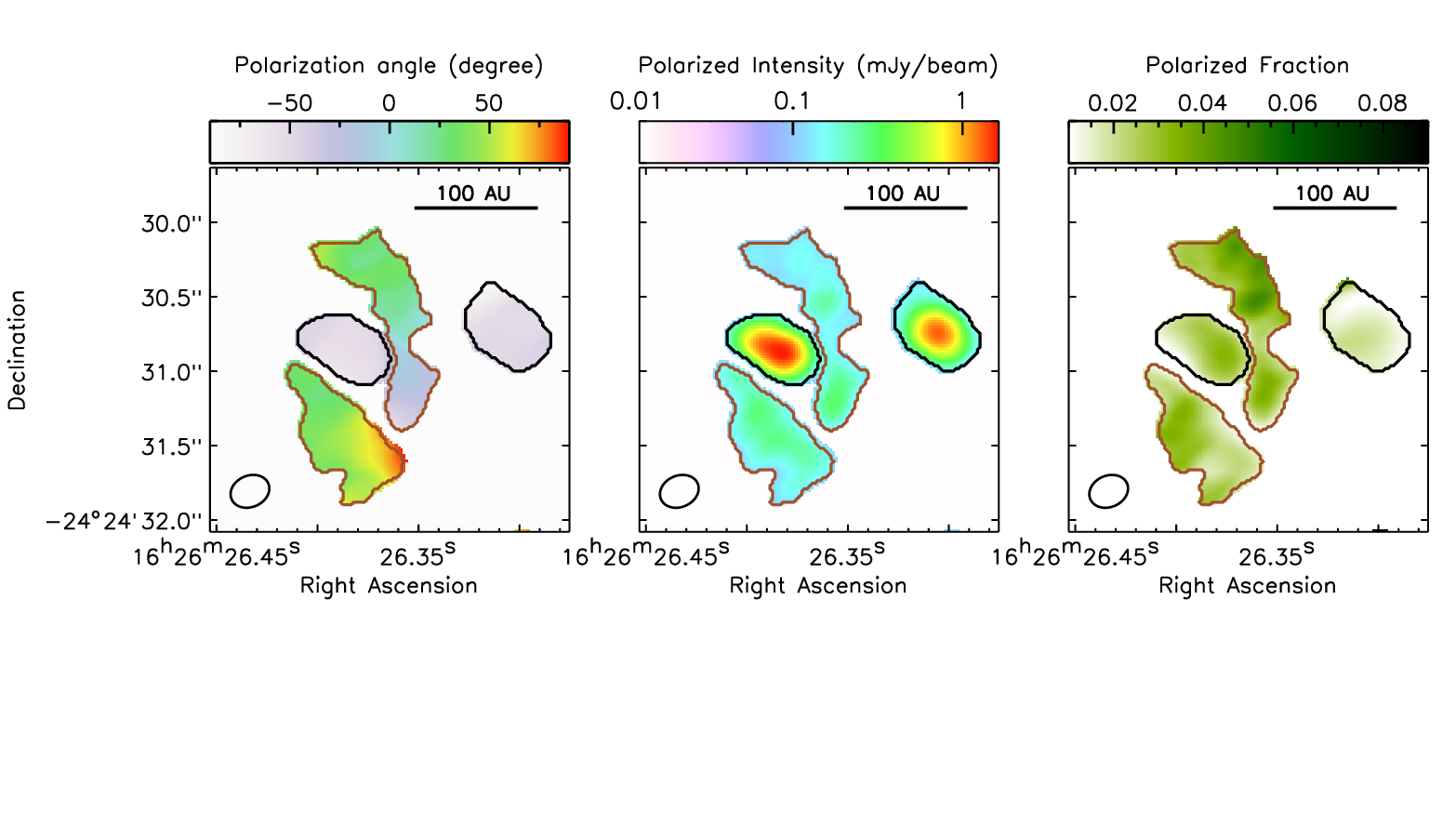}
\caption{The masks we use to identify the different components of VLA 1623.  Black contours show the mask for the optically thick inner protostellar disks.  Brown contours show the mask for the extended dust ring.  Background images show (left) polarization position angle, (middle) debiased polarized intensity, and (right) polarization fraction.  All images are limited to data with $\PI > 4\ \sPI$.    \label{masks}}
\end{figure*}

To study the mechanisms of polarization, we consider the inner protostellar disks and the extended dust ring as two separate subregions.  Figure \ref{masks} shows the outline of these two subregions in black and brown contours on maps of polarization angle and polarized intensity.   We produce these boundaries using 4 \sPI\ contours in polarized intensity with a slight modification based on the polarization angle map to separate the disk of VLA 1623-A from the extended dust ring.  We use these regions to produce the angle distributions in Figure \ref{angle_hist} and to discuss the polarization structure for the inner protostellar disks and extended dust ring in the next two sections.

\subsection{The Inner Protostellar Disks} \label{disk_section}

Table \ref{disk_prop} lists the positions, fluxes, semi-major axes ($a$), semi-minor axes ($b$), and position angles of the two protostellar disks (see Figure \ref{masks}) based on Gaussian fits to the Stokes I data.  We fit the disks using the \emph{imfit} task in MIRIAD \citep{Sault95}.   The values for semi-major axis, semi-minor axis, and position angle have been deconvolved with the beam, with errors estimated by the largest absolute difference between the best-fit deconvolved parameters and the range of deconvolved parameters when considering the errors in the observed Gaussian fit from \emph{imfit}.   These peak and flux errors correspond to statistical uncertainties, whereas the observation uncertainties are likely 10\%.  The disk position angle is measured from North to East.  We do not subtract a background level from these measurements. 

{\setlength{\extrarowheight}{0.8pt}%
\begin{table}[h!]
\caption{Results from Gaussian Fits to the Inner Protostellar Disks}\label{disk_prop}
\begin{tabular}{lll}
\hline\hline
				&	VLA 1623-A		& VLA 1623-B	\\
\hline
RA (J2000)		&	16:26:26.396		&  16:26:26.306 \\
Dec (J2000)		& 	-24:24:30.86		&  -24:24:30.72 \\
a$_{decon}$ (arcsec)			&	$0.368 \pm 0.041$		& $0.314 \pm 0.011$		\\
b$_{decon}$ (arcsec)			&	$0.188 \pm 0.041$		& $0.102 \pm 0.020$	\\
PA (degree)		&	$75.3 \pm 10.0$	& $42.4 \pm 2.7$	\\
peak (\mJybeam)	&	$60.6 \pm 3.6$		& $67.0 \pm 1.3$ 	\\
flux (mJy)			&	$141.8 \pm 5.9$	& $127.3 \pm 1.9$ 	\\
\hline
\end{tabular}
\end{table}
}

The polarization pattern of both disks shows very uniform morphologies with orientations of roughly $-50$\degree\ (see Figure \ref{angle_hist}) and median polarization fractions of 1.7\%.    This polarization morphology is aligned with the larger-scale collimated outflow (see Figure \ref{stokesI}) at a position angle of $-55$\degree\ \citep[e.g.,][]{Santangelo15}.  If the e-vectors trace grain alignment from an external magnetic field, then the magnetic field would be perpendicular to the axis of rotation as traced by the outflow.  Misaligned magnetic fields are significant in ideal magnetic hydrodynamical (MHD) simulations, because they mitigate magnetic braking  and allow large disks to form at early times in the star formation process \citep[e.g.,][]{HennebelleCiardi09, Machida11, Joos12}.  

\begin{figure}[h!]
\includegraphics[width=0.475\textwidth,trim=1pt 1pt 1pt 1pt,clip=true]{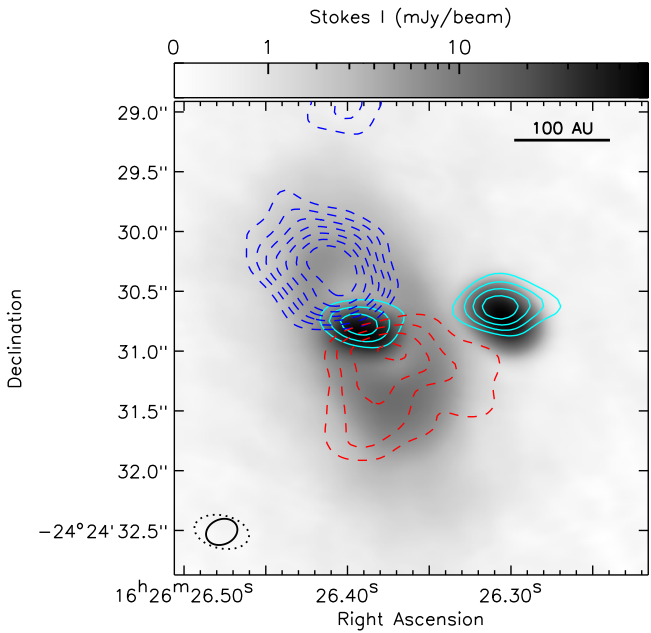}
\caption{Comparison of ALMA 1.3 mm continuum and ALMA Cycle 2 molecular line integrated intensities from project 2013.1.01004.S.  Background image shows our Stokes I dust continuum observations.  Cyan contours show \DCOp\ (3-2) absorption at levels of -10.5, -17.5, -24.5, and -31.5 \mJybeam\ \kms.  Red and blue dashed contours show red- and blue-shifted \COO\ (2-1) emission at levels of 35, 49, 63, 77, 91, 105, and 119 \mJybeam\ \kms.  The resolution of the Cycle 2 observations are given by the dotted ellipse in the lower-left corner.  The Cycle 2 observations are taken directly from the ALMA archive and are unpublished to date.  \label{dcop}}
\end{figure}

Polarization from magnetically-aligned dust grains is best detected when the dust emission is optically thin.  \citet{Yang17} showed that dust polarization from aligned dust grains decreases rapidly for large optical depths \citep[see Figure 3 in][]{Yang17}.   We can estimate the optical depth of the two disks using their spectral indices, $\alpha$ (e.g., with \textsf{nterms$=$2} in $clean$).  Both disks have $\alpha \approx 2$ (uncertainties $\lesssim 0.02$), which suggests that their dust emission is consistent with optically thick dust that is radiating like a pure blackbody.  The two protostellar disks show absorption features in unpublished high-resolution \CCO\ (2-1), \COO\ (2-1), and \DCOp\ (3-2) from ALMA Cycle 2 observations (project code 2013.1.01004.S).  Figure \ref{dcop} shows integrated intensity of the high density tracer, \DCOp\ (3-2), in cyan contours at levels of -10.5, -17.5, -24.5, and -31.5 \mJybeam\ \kms.  On larger scales, \DCOp\ (3-2) is seen in emission outside of the disk \citep[e.g.,][]{Murillo15}, indicating that these disks are consistent with optically thick structures.

For $\tau > 1$, polarization from self-scattering is expected to dominate over polarization from grain alignment \citep{Yang17}.  A number of studies have examined models of dust scattering with a variety of disk properties, such as gaps, dust settling, asymmetries, and different inclinations \citep[e.g.,][]{Kataoka15, Kataoka16, Pohl16, Yang16, Yang16_iras4a,Yang17}.  These models tend to show distinct polarization morphologies based on variations in the anisotropic radiation field.  For example, face-on disks generally produce self-scattering polarization patterns with azimuthal morphologies, whereas highly inclined disks have uniform e-vectors that are primarily aligned with the minor axis of the disk \citep[e.g.][]{Pohl16, Yang16}.   Polarization fractions from these dust scattering models are generally a few percent.

Most of the aforementioned studies use disk models with optical depths of $\tau \lesssim 1$, whereas the inner protostellar disks around VLA 1623-A and VLA 1623-B appear optically thick.  \citet{Yang17} modeled self-scattering toward an inclined (45\degree) optically thick disk and found polarization morphologies along the direction of the minor axis similar to the $\tau \approx 1$ disks.  \citet{Yang17} also found that the viewing angle affected the plane-of-sky polarized intensities.  For an inclined, optically thick disk, the near-side of the disk should have a higher fraction of polarized light than the far-side.  This asymmetry should manifest as higher polarized intensities and fractions on one side of the disk minor axis.

Qualitatively, VLA 1623-A and VLA 1623-B have polarization morphologies and fractions that are consistent with the optically thick model from \citet{Yang17}.  First, the disks have median polarization fractions of 1.7\%, which is consistent with dust self-scattering models.  Second, they have uniform polarization orientations within 35$\degree$ (VLA 1623-A) and 5$\degree$ (VLA 1623-B) of their minor axes.   Both disks are more inclined than the model presented in \citet{Yang17}, although disks with steeper inclinations should still have their e-vectors mostly aligned with their minor axes.  We estimate inclination angles of $i=60$\degree\ for VLA 1623-A and $i=70$\degree\ for VLA 1623-B,  assuming $\cos{i} = b/a$ (see Table \ref{disk_prop}).   We also see evidence of asymmetric polarized intensities in VLA 1623-A (see Figure \ref{masks}), although the asymmetry is present along both the major and minor axes.  The polarized intensities of VLA 1623-B are more Gaussian-like, but we may lack the resolution to definitively trace any asymmetries in its disk.   

Overall, the dust polarization seen toward the inner protostellar disks of both VLA 1623-A and VLA 1623-B are more consistent with signatures of self-scattering than grain alignment.  These two disks join a growing list of systems with strong evidence of dust scattering from high-resolution observations; HD 142527 \citep{Kataoka16}, Cepheus A HW2 \citep{FernandezLopez16}, HL Tau \citep{Stephens14, Stephens17}, HH 212 / HH 111 \citep{Lee18}, and several embedded sources in Perseus \citep{Cox18}.  We discuss the implications of dust scattering in Section \ref{scattering}.

\subsection{The Extended Dust Ring} \label{ring_section}

\begin{figure}[h!]
\includegraphics[angle=-90,width=0.475\textwidth,trim=1pt 1pt 1pt 1pt,clip=true]{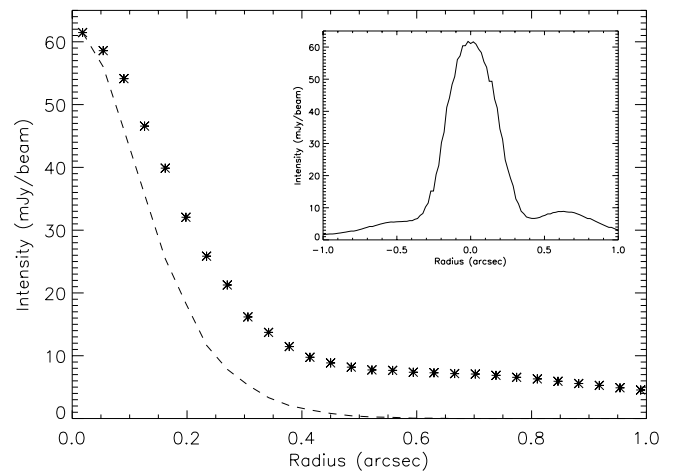}
\caption{Azimuthally-averaged profile of VLA 1623-A.  The profile is centered on the peak of emission for VLA 1623-A and uses elliptical apertures with a position angle of 30\degree\ and axis ratio of $\approx 0.58$ to match the observed ring shape.  The radii correspond to the radial extent along the long axis.  The dashed curve shows the radial profile of the beam, scaled to match the peak of VLA 1623-A.  The inset shows an intensity slice along the long axis of VLA 1623-A, where the gap-like feature is more prominent.  \label{intensity_profile}}
\end{figure}

We see an extended dust ring around VLA 1623-A.  Figure \ref{masks} outlines the boundary we used for its polarized emission.  Hereafter, we use the term ``ring'' to describe this structure because it appears to have a resolved, ring-like morphology (e.g., see Figures \ref{stokesI} and \ref{dcop}).  Figure \ref{intensity_profile} shows the azimuthally-averaged intensity profile of VLA 1623-A using elliptical apertures that match the observed shape of the ring (see below).   The intensity profile, however, does not show dips as we would expect for a true ring structure with a gap in the center.  Instead, we cannot differentiate the intensity profile from the average profile of a disk.  Nevertheless, the inset in Figure \ref{intensity_profile} shows an intensity slice through the major axis of the ring where a clear dip is seen at a radial extent of $\approx 0.4$\arcsec\ from VLA 1623-A.  This dip suggests that a gap may be present as expected given the ring-like morphology seen in Stokes I.  A gap between VLA 1623-A and the ring will be discussed in detail by Harris et al. (submitted).   

The intensity structure of the ring is asymmetric.  The peak emission in the southern portion appears brighter than the northern portion by a factor of three.  This asymmetry may be caused by differences in temperature, mass, or dust opacity.  At this time, we cannot determine the origin of the asymmetry.   We estimate a size of $1.44\arcsec \times 0.83\arcsec$ and a position angle of 30\degree\ by eye using an ellipse that approximates the center of the ring.   Assuming this ellipse is an inclined circle, we find an inclination of 55\degree, which is similar to the inclination of the inner protostellar disk.  The inner disk, however, has a different position angle (75\degree) from the ring, which makes their relationship more complex.  Based on higher resolution Band 7 observations of VLA 1623 (Harris et al., in preparation), VLA 1623-A may contain an unresolved, tight binary system that could be related to the offset in position angle. 

The ring also coincides with red- and blue-shifted \COO\ (2-1) line emission from earlier ALMA observations.  Figure \ref{dcop} shows contours of red- and blue-shifted \COO\ (2-1) integrated intensities from ALMA Cycle 2 observations on our map of the ring.   \citet{Murillo13} used Cycle 0 \COO\ (2-1) emission to identify a pure Keplerian disk with a (minimum) radius of 50 au and used models to estimate the Keplerian disk extends to at least 150 au.  Using the \COO\ (2-1) kinematics, \citet{Murillo13} estimated an inclination of 55\degree\ assuming circular symmetry.  This inclination matches our estimate based on the geometry of the ring. 

Figure \ref{polarization} shows an azimuthal polarization morphology over roughly three-quarters of the ring with a typical polarization fraction of 2.4\%\ \citep[see also BHB07-11 in][]{Alves18}.   The north-east region of the ring is not detected in polarization.   For this non-detection, we estimate a 4-$\sigma$ upper limit of 2\%, which is comparable to the typical polarization fraction for the entire ring.  Even if we lower our threshold to 3-$\sigma$ (upper limit polarization is 1.5\%), we do not detect any polarization in the north-east quadrant (see also, Figure \ref{stokesQU}).  Thus, this non-detection is significant and indicates that the north-east quadrant of the ring has genuinely lower polarization than the rest of the ring.

Models of dust scattering primarily focus on disks rather than rings.  Nevertheless, we can approximate the ring as the outer portion of a disk with a large gap between in the inner and outer extents.  \citet{Pohl16} modeled dust self-scattering in disks with gaps and found that the outer ring shows an azimuthal polarization morphology when viewed face-on and a morphology mostly aligned with the minor axis when viewed at a steep inclination.   This result is similar to those models with continuous disks \citep[e.g.,][]{Kataoka15, Kataoka16hd, Yang16, Yang17}.  At an inclination of $\sim 55$\degree, the model disk with a large gap should have e-vectors that are skewed toward the minor axis \citep{Pohl16}.   Since we see an azimuthal polarization pattern, we do not attribute the observed polarization in the ring to pure self-scattering.  

\begin{figure}[h!]
\includegraphics[width=0.475\textwidth,trim=1pt 1pt 1pt 1pt,clip=true]{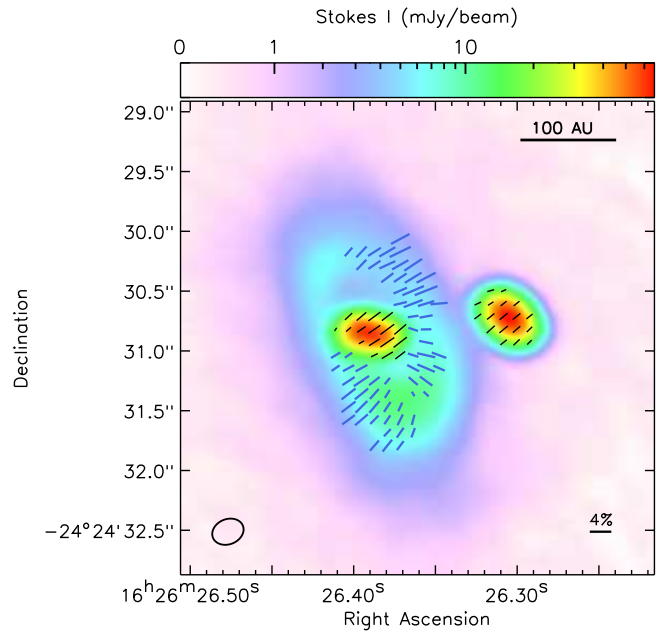}
\caption{Magnetic field orientation for the ring around VLA 1623-A.  The blue b-vectors toward the ring have been rotated by 90\degree.  The black e-vectors toward the inner protostellar disks are not rotated.   Note that the vector lengths are shortened relative to Figure \ref{polarization} to highlight the b-vector morphology.   \label{bfields}}
\end{figure}

We consider instead polarization from grain alignment due to magnetic fields or an anisotropic radiation field.  For magnetic grain alignment, we rotate the e-vectors toward the ring by 90\degree\ (hereafter, called b-vectors) to infer the magnetic field direction.  Figure \ref{bfields} shows the b-vectors associated with the ring in blue.  (Note that the black vectors toward the inner protostellar disks are unrotated e-vectors.)   The b-vectors appear radial, which is inconsistent with a pure toroidal or poloidal magnetic field structure.  Instead, a radial magnetic field may correspond to more complicated field geometries like hourglass or quadrupole shapes \citep{Frau11, Reissl14}.   Hourglass morphologies in particular are expected for contracting clouds with flux-frozen magnetic fields \citep[e.g.,][]{MestelStrittmatter67, Mouschovias76, GalliShu93}.  

Figure \ref{bfields} shows hints of curvature in the b-vectors that could represent an hourglass morphology.  We may not cover a wide enough range of spatial scales to fully trace the pinching pattern from a flux-frozen field, however.  Theoretical predictions suggest the hourglass morphology should trace scales of hundreds to a few thousand AU \citep[e.g.,][]{GalliShu93, Allen03, Frau11, Kataoka12}, which matches the hourglass shapes detected in observations \citep[e.g.,][]{Girart06, Rao09, Stephens13}.   We estimate a maximum recoverable scale of $2.6$\arcsec\ using the fifth percentile\footnote{This is the same percentile used in the ALMA Technical Guide when outlining the maximum recoverable scale for each array configuration.} baseline (102.275 m) and an average elevation of 80\degree.  At a distance of 120 pc, this maximum recoverable scale spans only 310 au.   Additional short spacings may help qualitatively trace the possible hourglass structure.  We discuss an hourglass field geometry in more detail in Section \ref{Bfield_section}.

For grain alignment from an anisotropic radiation field \citep{LazarianHoang07, Tazaki17}, dust grains precess due to torques from the radiation field and align with their long axes perpendicular to it.   For most disks, the radiation field is generally radial (e.g., the gradient of radiation decreases outward from the central star) and dust grains in this radiation field will produce an azimuthal polarization pattern at (sub)millimeter wavelengths. 

Qualitatively, the polarization morphology seen in the ring shows an azimuthal pattern (see Figure \ref{polarization}), as expected for dust grains aligned by the anisotropic radiation field.  Recent observations of HL Tau similarly show azimuthal polarization at 3 mm \citep{Kataoka17} and at 1.3 mm \citep{Stephens17}, which have been attributed to grain alignment from the anisotropic radiation field.  The 1.3 mm observations in particular show a mix between dust scattering toward the optically thick center of the HL Tau disk and radiative grain alignment toward the optically thin outer radii of the disk.  The polarization data for VLA 1623-A could represent the same situation.  We discuss grain alignment from an anisotropic radiation field in more detail in Section \ref{Rfield_section}.

\subsection{Comparison to Previous Polarization Observations}\label{compare_obs}

Several previous studies have used single-dish telescopes to map dust polarization across $\rho$ Oph A, which includes VLA 1623.  In the first such study, \citet{Holland96} detected a single vector with a position angle of $-70$\degree\ toward VLA 1623 at 800 \um\ with the UKT14 detector at the James Clerk Maxwell Telescope (JCMT).  Successive single-dish dust polarization maps of $\rho$ Oph A yielded more vectors at 350 \um\ with Hertz at the Caltech Submillimeter Observatory \citep{Dotson10} and at 850 \um\ with SCUPOL \citep{Matthews09} and POL-2 (Kwon et al., submitted) at the JCMT.   These studies showed that the polarization structure toward VLA 1623 was relatively uniform, with polarization position angles of roughly $-20$\degree\ to $-30$\degree, and polarization fractions of $\PF < 2$\%.  

Using CARMA, \citep{Hull14} measured dust polarization toward VLA 1623 at 1.3 mm and $\sim 3.3$\arcsec\ resolution.  These data yielded mostly uniform e-vectors with orientation of roughly $-60$\degree\ across both VLA 1623-A and VLA 1623-B (the two protostars were unresolved).  The CARMA data also appears to bend to a position angle of roughly 15\degree\ in a region 6\arcsec\ north-east of VLA 1623-A.  Nevertheless, there is only one robust ($\PI > 3.5\ \sPI$) vector in this north-east region, so this change in polarization morphology is still ambiguous.  VLA 1623 was also observed in dust polarization at 870 \um\ with the SMA, but those data lacked the sensitivity to detect polarization robustly \citep{MurilloLai13}.

There is no indication of the azimuthal polarization structure we detect toward the ring in the CARMA observations.  Since the polarization structure of the ring is roughly symmetric, the e-vectors may cancel out if they are unresolved in the larger CARMA beam.   As a test, we convolve our ALMA observations to the same resolution ($4.3\arcsec\ \times 2.5\arcsec$, $\theta = 7.6$\degree) as the VLA 1623 data presented in \citet{Hull14} and remeasure the e-vector position angles.  From our smoothed data, we find a typical position angle of -57\degree, which matches \citet{Hull14}.   Thus, the CARMA observations may be primarily tracing dust polarization from self-scattering in the inner protostellar disks even though the disks themselves are unresolved.


\section{Discussion} \label{discussion}

\subsection{Magnetic Field Alignment} \label{Bfield_section}

Models of magnetic fields that thread dense cores usually assume initially uniform morphologies that are flux frozen.  When the cores condenses, the field lines are dragged inward and produce an ``hourglass'' shape.  The inward pinch of the field lines enhances the magnetic pressure and causes an unbalanced magnetic force, which concentrates material toward the equatorial plane \citep[e.g.,][]{Mestel66, MestelStrittmatter67, GalliShu93}.   If the magnetic field is strong enough, the collapsing material will form a flattened structure often called a ``pseudodisk'', which differs from a rotationally-supported Keplerian disk because it is out of equilibrium, although a pseudodisk itself can rotate \citep[e.g.,][]{Masson16}.   They are $\sim 100-1000$ au in size according to models and simulations \citep[e.g.,][]{GalliShu93_pseudosize, Kataoka12, EwertowskiBasu13}, but none have been definitively identified from observations of low-mass regions to date.

In Section \ref{ring_section}, we show the inferred plane-of-sky magnetic field morphology associated with the ring of dust around VLA 1623-A.  This morphology shows a slightly-distorted radial structure that is inconsistent with a toroidal magnetic field, which should be predominantly spiral in structure during the accretion phase after the protostar forms \citep[e.g.,][]{Tomisaka11, Kataoka12}.  Instead, the dust polarization toward the ring appears more analogous to an hourglass magnetic field.  

To represent this hourglass morphology, we use a simple model of a flux-frozen magnetic field in a centrally condensed oblate spheroid that is inclined with respect to the line of sight.  The model is only a first step toward a full description, which should include both disk rotation and the dense core environment.  Nonetheless, the model gives a reasonable fit to the observed polarization directions and serves as a first look at the magnetic field structure of VLA 1623-A.

\subsubsection{Model Description}\label{about_model}

We assume a weak magnetic field with a magnetic energy density that is small compared to self-gravity.  The magnetic flux is frozen along the short axis of an oblate Plummer spheroid with peak density $n_0$, scale length $r_0$, and index $p=2$ \citep{Plummer11, Arzoumanian11, Myers17}.  We assume this spheroid condensed from a larger cloud of uniform density $n_u$ and uniform field strength $B_u$, while conserving mass, flux, and shape.  

We describe the magnetic flux, $\Phi$, as a series of concentric flux tubes expressed analytically in terms of $n_0$, $n_u$, and $B_u$ \citep{Mestel66, MestelStrittmatter67}.  The ``flux profile'' of a given flux tube can be written in terms of the dimensionless flux $f = \Phi/(\pi B_ur_0^2)$ and dimensionless coordinates $\xi = x/r_0$, $\eta = y/r_0$, and $\zeta = z/r_0$ as
\begin{equation}
f(\xi,\eta,\zeta) = \xi_c^2\left[1+\left(\frac{3\nu_0}{\omega^2}\right)\left(1-\frac{\tan^{-1}\omega}{\omega}\right)\right]^{2/3} , \label{flux_eq}
\end{equation}
where $\xi_c = \sqrt{\xi^2 + \eta^2}$ is the cylindrical radius, $\nu_0 = n_0/n_u$ is the ratio of peak to background density, and the dimensionless radius $\omega$ is given by
\begin{equation}
\omega^2 = (\xi^2+\eta^2)/A^2 + \zeta^2 , \label{radiusEq}
\end{equation}
for an aspect ratio of $A = 10$.  We assume the spheroid has initial orientation with its symmetry axis (short axis) along the $z$-direction in the plane of sky (e.g., $\eta = 0$).  Then, $\xi_c = \xi$ and $\omega^2 = \xi^2/A^2 + \zeta^2$.  

With this orientation, the spheroid will appear as a highly flattened ellipse with an aspect ratio of 10.  Since we view the ring as an ellipse with an aspect ratio of 1.7,  we incline the spheroid and its associated field lines through a polar angle of $\theta=54\degree$ from the plane of the sky toward the line of sight.  The field line coordinates in Equations \ref{flux_eq} and \ref{radiusEq} are then transformed to give a tilted hourglass shape.  Further details of this procedure and these models are described in a forthcoming publication (Myers et al. 2018, in preparation). We note that our model polarization pattern agrees well with the numerical simulations from \citet[][see their Figures 6 and 7]{Kataoka12}.  

\subsubsection{Model Results}\label{model_results}

To produce a realistic model, we need to estimate the peak density and scale height of the oblate spheroid.  We approximate the peak density by scaling the mean density of the ring by the ratio of peak density to mean density in a critical Bonnor-Ebert sphere.  We determine the mean density of the ring from estimates of its total mass and volume.  First, we subtract the the best-fit Gaussians for VLA 1623-A and VLA 1623-B (see Table \ref{disk_prop}) from our Stokes I continuum observations using \emph{imfit} in MIRIAD.  Second, we use this source-subtracted image to measure the 1.3 mm flux of the ring, $S_{1.3} = 0.6$ Jy. Third, we calculate a total mass for the ring $M = S_{1.3}d^2/B(T)\kappa = 0.1\ \Msun$, where $d$ is the distance to Ophiuchus, $B(T)$ the the blackbody function, and $\kappa$ is the dust opacity per unit dust and gas mass.  We assume $T = 20$ K and $\kappa = 0.024$ \cmg\ at 1.3 mm \citep{Andrews09}.  Finally, we divide the estimated mass in the ring by the volume of the oblate spheroid to get a mean density of $\sim 1 \times 10^{10}$ \vol\ and an adopted peak density of $\sim 6 \times 10^{10}$ \vol.   For the scale length, we adopt a size of $r_0 = \sigma/\sqrt{4\pi G\mu_{H}m_Hn_0} = 40$ au, assuming $\sigma$ is the thermal velocity dispersion for hydrogen gas at a temperature of 20 K, $G$ is the gravitational constant, $\mu_{H} = 2.33$ is the mean molecular mass per free particle, $m_H$ is the atomic hydrogen mass, and $n_0$ is our estimated peak density.  

\begin{figure}[h!]
\includegraphics[width=0.475\textwidth,angle=-90,trim=1pt 1pt 1pt 1pt,clip=true]{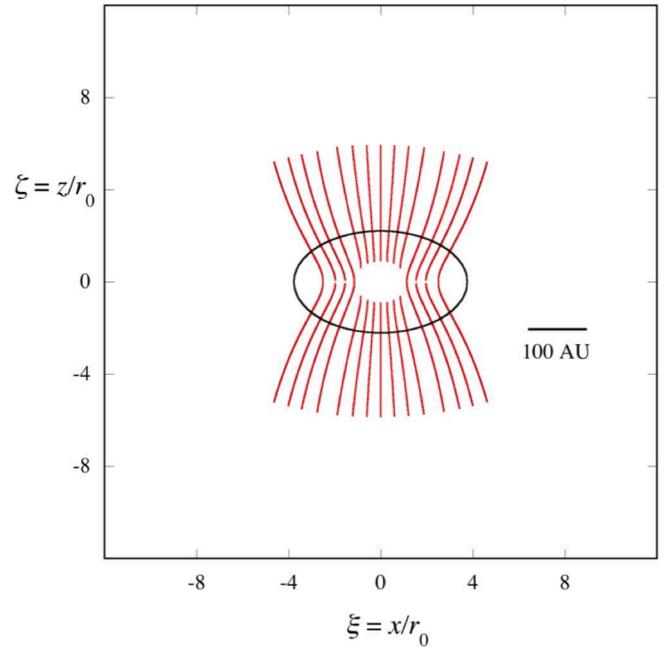}
\caption{Model magnetic field for an oblate spheroid inclined by 54\degree.  The red solid lines show the plane-of-sky magnetic field structure assuming the flux pattern is comprised of four concentric flux tubes.  The black ellipse shows the outline of the model spheroid.    \label{model}}
\end{figure}

We qualitatively match our model by eye to the observations.  Figure \ref{model} shows our matched magnetic field structure in the plane of the sky for the inclined oblate spheroid.  The red curves show the field lines and the black ellipse outlines the plane-of-sky shape for the inclined spheroid.  The flux tubes are selected to give similar spacings as the observed polarized data (e.g., following Nyquist sampling within the observed beam).    We also adopt a ratio of peak density to background density of $\nu_0 = 60$ to approximately match the observed polarization directions.   As expected for a flux-frozen magnetic field, we find a pinched morphology.   Since we assume a uniform background for our model, this density ratio is highly idealized and not a realistic description of the true density contrast between the disk and dense core environment.   The model should only be applied to the inner regions of VLA 1623 associated with the scales where we have ALMA observations.   The simple model is completely scalable, however, and can be adjusted to a different density ratio when more data are made available.

\begin{figure}[h!]
\includegraphics[width=0.455\textwidth,angle=-90,trim=1pt 1pt 1pt 1pt,clip=true]{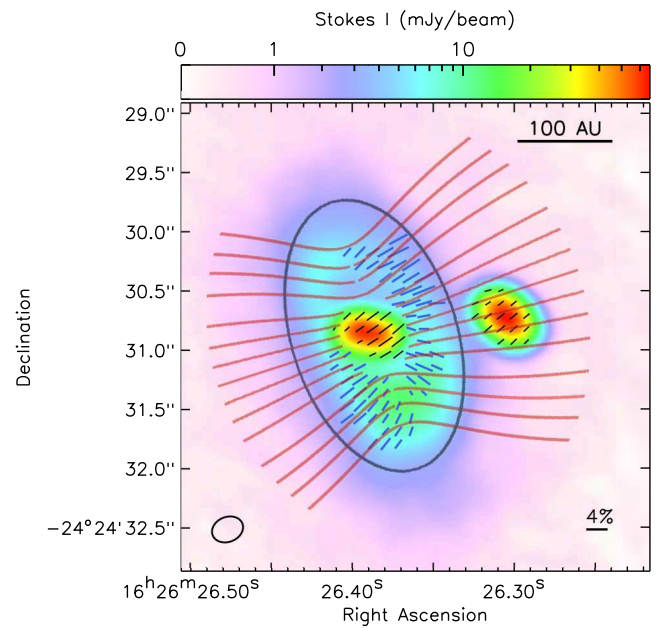}
\caption{Overlay of our model magnetic field with our observations of VLA 1623.  The background image is the same as \ref{bfields}, where black vectors are e-vectors and blue vectors are b-vectors.  The red solid lines are the same as in Figure \ref{model}, but have been rotated to match the position angle of the ring.  The black ellipse shows the outline of the model spheroid.   \label{model+obs}}
\end{figure}

Figure \ref{model+obs} shows an overlay of the model magnetic field structure (Figure \ref{model}) rotated by a position angle of 30$\degree$ with our observed b-vectors toward the ring (Figure \ref{bfields}).  We find good by eye agreement between the observed and modeled plane-of-sky magnetic field orientations, particularly in the north-west quadrant and south-east quadrant.  There are noticeable departures in the north-east and south-west quadrants, however. The former is where we detect no dust polarization.  These deviations may be linked to our assumption of axisymmetry.   The intensity distribution for the ring is clearly asymmetric, which suggests its mass distribution may also be asymmetric.   

Nevertheless, even with our simple assumption of axisymmetry, the model approximates the inferred b-vector morphology.   This agreement implies that the emission around VLA 1623-A corresponds to a disk with evidence of magnetization and that the magnetic field axis is primarily aligned with the direction of the large-scale, collimated outflow.  Under the assumption of ideal MHD, this field orientation should suppress disk formation \citep[e.g.,][]{HennebelleCiardi09, Machida11}.  Since \citet{Murillo13} identified Keplerian rotation out to 180 au toward VLA 1623-A and we see a resolved compact disk-like structure toward VLA 1623-B, ideal MHD may not be applicable to full radial extent of VLA 1623.  Recent observations of the Class 0 source B 335 from ALMA \citep{Maury18} are well-matched with models of non-ideal MHD collapse.  Moreover, non-ideal MHD simulations readily produce disk structures with Keplerian rotation profiles, even in cases with relatively strong magnetic fields \citep[e.g.,][]{Tomida15, Masson16, Hennebelle16, Vaytet18}.   Thus, VLA 1623-A is an excellent candidate to study the effects of non-ideal MHD in the formation and evolution of a large disk $\sim 200$ au in diameter.

As previously stated, our simple model excludes rotation.  VLA 1623-A has a large rotating disk that follows a Keplerian profile \citep{Murillo13} and rotation is expected to draw a vertical magnetic field into a toroidal field on $< 100$ au scales \citep{Tomisaka11, Kataoka12}.  Nevertheless, other models suggest that the toroidal component is either localized or weak.  \citet{Masson16} produced an 80 au disk with Keplerian-like rotation in non-ideal MHD simulations and found that it had a weak toroidal component and strong poloidal component in spite of its rotation \citep[see also,][]{Tomida15, Zhao18}.   In non-ideal MHD simulations at the formation of the second core, \citet{Vaytet18} produced a strong toroidal component only in the vicinity of the second core ($r \lesssim 1$ au), which is on scales too small for us to probe.  Thus, it is unclear how rotation influences magnetic field directions and the subsequent polarization on the scales we can observe.    We require more detailed models that include rotation and non-ideal MHD effects to fully represent the observed polarization structure in VLA 1623-A \citep[e.g.,][]{Maury18}.

\subsection{Radiation Field Alignment} \label{Rfield_section}

Dust polarization can also arise from grain alignment due to anisotropic radiative torques from the radiation field itself.  Based on theoretical studies, large ($\gtrsim$ 100 \um) dust grains are more efficiently aligned by the radiation field than magnetic fields \citep[e.g., see Figure 5 in][]{Tazaki17}, although the precession rate for the latter depends on the amount of paramagnetic material in the dust grains \citep{HoangLazarian16}.   Observations of disks at millimeter wavelengths trace primarily these large dust grains, so radiative grain alignment should expected \citep[e.g.,][]{Kataoka17, Stephens17}.  

\begin{figure}[h!]
\includegraphics[width=0.475\textwidth,trim=1pt 1pt 1pt 1pt,clip=true]{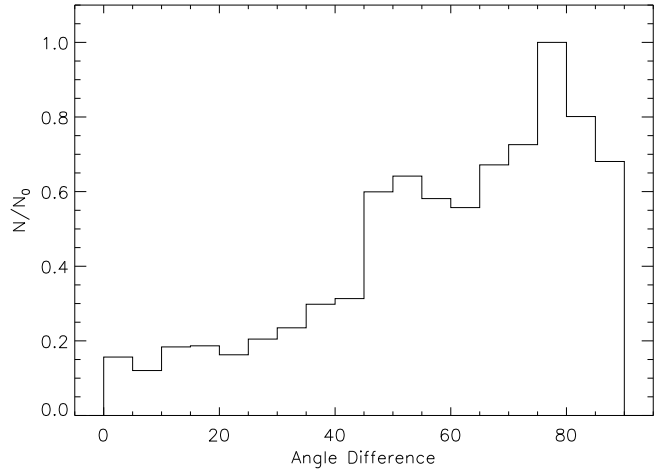}
\caption{Normalized distribution of relative angles between the Stokes I intensity gradient and the polarization position angles for the ring around VLA 1623-A.  We consider only those pixels $I > 3\ \sigma_I$ and $\PI > 4\ \sPI$ associated with the ring (see Figure \ref{masks}).  \label{ang_diff}}
\end{figure}

A simple test for radiative grain alignment is to compare the direction of polarization to the gradient of radiation.  Grains with their long axes perpendicular to the radiation field will primarily have 90\degree\ differences between Stokes I intensity gradient and the polarization position angles.  Figure \ref{ang_diff} shows these relative angles for the ring only.   The distribution peaks at roughly 80\degree\ and decreases sharply at relative angles $\lesssim 45$\degree, suggesting that the Stokes I intensity gradient is primarily perpendicular to the polarization pattern.  Thus, the observations are also consistent with grain alignment from an anisotropic radiation field.

Figure \ref{ang_diff}, however, is also relatively broad.  A large fraction of the ring has relative angles of $50-60$\degree, which indicates that the polarization is neither strongly aligned nor strongly perpendicular to the direction of the radiation field.  These differences could indicate that more than one mechanism is affecting the polarization pattern (e.g., magnetic fields, see Section \ref{Bfield_section}).  Alternatively, the broadness could mean that grain alignment from the anisotropic radiation field is less efficient.  This mechanism is expected to affect mainly large dust grains, and VLA 1623 (a Class 0 system) may not have had enough time to produce a large population of these grains although low measurements of the dust emissivity index, $\beta$, indicate that some level of grain growth should be expected in Class 0 systems \citep[e.g.,][]{Kwon09, Chiang12}.

If the polarization pattern in the ring is due to radiative grain alignment, then we can expect large dust grains in the ring as small grains are more affected by gaseous damping and will not align efficiently.  \citet{Tazaki17} modeled both magnetic and radiative grain alignment in a protoplanetary disk.  Assuming thermal dust emission of 20 K in the disk midplane ($\lambda \approx 140$ \um), they found that the radiative torques are most efficient for grain sizes $\gtrsim 80$ \um\ at radial distance of 50 au.  If the gas surface density is lower then gaseous damping will be weaker.  Thus, radiative torques may be more efficient for smaller grains in the surface layers of disks (e.g., if the disks have some thickness) or at larger radial extents (e.g., 100 au).  

Models of dust growth in disks \citep[e.g.,][]{Brauer08, Birnstiel10} show that the time scale for grain growth to $\sim 100$ \um\ sizes depends significantly on the local density of the disk, which is strongly a function of radius.  For example, millimeter-sized  dust grains are expected to grow within the Class 0 lifetime \citep[e.g., $\sim 0.2$ Myr;][]{Dunham15} in the inner few au of disks ($\Sigma \sim 10^2$ g cm$^{-1}$), whereas at distances of 100 au ($\Sigma \sim 10^{-1}$ g cm$^{-1}$), grains are generally $< 100$ \um\ in size \citep[e.g.,][]{Brauer08, Birnstiel10}.  

The ring spans a range of radii from $\sim 50-90$ au.  If the polarization from the ring is produced by radiative grain alignment, then we can expect dust grain sizes comparable to or a bit less than 80 \um.  These grain sizes may fit with grain growth models, although we need estimates of size distributions on more intermediate extents (e.g., 75 au).   Alternatively, if the grains are still relatively small ($\sim 10$ \um) in the ring, then magnetic grain alignment may be more efficient \citep{Tazaki17}.

\subsection{Dust Scattering} \label{scattering}

Polarized dust emission can arise from self-scattering off dust grains.  This effect has been readily utilized to study debris disks around stars at infrared wavelengths \citep[e.g.,][]{Perrin15}, but was historically overlooked at (sub)millimeter wavelengths due to the assumption of $\tau \ll 1$.    \citet{Kataoka15} first demonstrated that self-scattering in protostellar and protoplanetary disks can produce polarization signatures that are easily measurable with ALMA  if the dust grains have reach $\sim 100$ \um\ sizes \citep[e.g., see also][]{Pohl16, Yang16, Yang17}.  Such large grains are typically not detected in diffuse molecular clouds, although they are expected at small radial extents in protostellar and protoplanetary disks \citep[e.g.,][]{DAlessio01, Brauer08, Birnstiel10}. 

We see evidence of polarized self-scattering toward the VLA 1623-A and VLA 1623-B inner protostellar disks (see Section \ref{disk_section}).    This result indicates that the inner protostellar disks may have already formed large dust grains of a few hundred microns in size.  In particular, \citet{Kataoka15} found that polarization from scattering is most efficient for dust grains with sizes of $\sim \lambda/2\pi$ \citep{Kataoka15}, which corresponds to $\approx 200$ \um\ for $\lambda = 1.3$ mm.  This result matches well the expected timescales for grain growth in the inner $\sim 30$ au of disks, where the surface densities are high and grains of $\sim 100$ \um\ sizes can form in $\sim 0.2$ Myr \citep[e.g.,][]{Brauer08, Birnstiel10}.  

Dust scattering observations may also be excellent tools to identify true protostellar disks.  \citet{Maury12} proposed that VLA 1623-B is a knot produced by the collimated outflow rather than a protostellar source itself.  The polarization pattern for VLA 1623-B is consistent with dust self-scattering; e.g., the polarization position angles are both uniform and aligned within 5\degree\ of its minor axis.  We would not expect this type of dust polarization from an outflow knot, as these structures would not have the large dust grains needed to efficiently self scatter.  Thus, we can conclude that we are most likely tracing dense disks around VLA 1623-B \citep[e.g., see also,][]{Murillo13}.

VLA 1623-A, however, shows a clear transition in dust polarization that we attribute to different mechanisms.  We see evidence of dust scattering toward the inner disk and evidence of grain alignment from either the magnetic field or the anisotropic radiation field toward the ring.  The transition occurs near an intensity level of $I \approx 20$ \mJybeam\ for the inner disks.  Assuming a dust temperature of 20 K and a dust opacity of 0.024 \cmg\ \citep{Andrews09}, the transition point corresponds to a column density of $\NHH \approx 5 \times 10^{24}$ \cden.  We caution that this column density is likely a lower limit as the inner disks appear to be optically thick.

\subsection{Mechanical Grain Alignment} \label{other_mech}

We find that our polarization observations are best explained by dust scattering in the dense, inner, protostellar disks and either magnetic grain alignment or radiative grain alignment in the ring.  Here we discuss an alternative process that can cause dust polarization and the impact it may have on our observations.  This mechanism is mechanical grain alignment.

Mechanical grain alignment was first described by \citet{Gold52}.  It arises from interactions between dust grains and gas flows, where collisions between the dust and gas transfer angular momentum to the dust across the direction of motion.  This transfer of angular momentum will preferentially align the dust with their long axes parallel to the gas flow \citep[e.g.][]{Purcell69, Andersson15}.  In general, mechanical grain alignment is most efficient for supersonic gas flows \citep{Lazarian97} such as those associated with outflows \citep[e.g.,][]{Rao98,Cortes06}.   VLA 1623 has a strong, collimated outflow \citep{Andre90}, which could be the source for mechanical grain alignment.  

For the ring, we see azimuthal dust polarization (see Figure \ref{polarization}), which is inconsistent with mechanical grain alignment from the large-scale outflow.  The polarization angles for the inner protostellar disks, however, are consistent with the outflow direction.  Nevertheless, we do expect the outflow to strongly impact them  First, \citep{Murillo13} found evidence of Keplerian rotation on scales of $\sim 180$ au.  This rotation implies that VLA 1623-A has kinematics independent of the outflow.  Second, millimeter continuum emission generally originates toward the midplane of inner protostellar disks \citep{Dullemond07} and not the upper layers of the disk where the outflow may be launched \citep{Pudritz07}.  Thus, mechanical grain alignment from the outflow is unlikely to impact our observations.

Alternatively, the dust grains could be aligned by other kinematic processes.  Recent studies have shown that dust grains with highly irregular shapes can aligned efficiently with subsonic gas flows  \citep[e.g.,][]{LazarianHoang07, DasWeingartner16, Hoang18}.  Thus, we may need to consider mechanical grain alignment from gas infalling onto the protostars or the Keplerian motions of the disks themselves.   At this time, we do not have evidence of infalling gas flows onto the protostars.  Grain alignment from Keplerian motions, however, is possible and would produce an azimuthal pattern such as what is seen in the ring.  Keplerian motions at large angular extents are relatively slow, so it is unclear whether or not they can efficient align grains.    So at this time, we do not consider mechanical grain alignment to be a likely cause of the polarization detected in VLA 1623.

\subsection{Proper Motion of VLA 1623} \label{offset_section}

\begin{figure}[h!]
\includegraphics[width=0.475\textwidth,trim=1pt 1pt 1pt 1pt,clip=true]{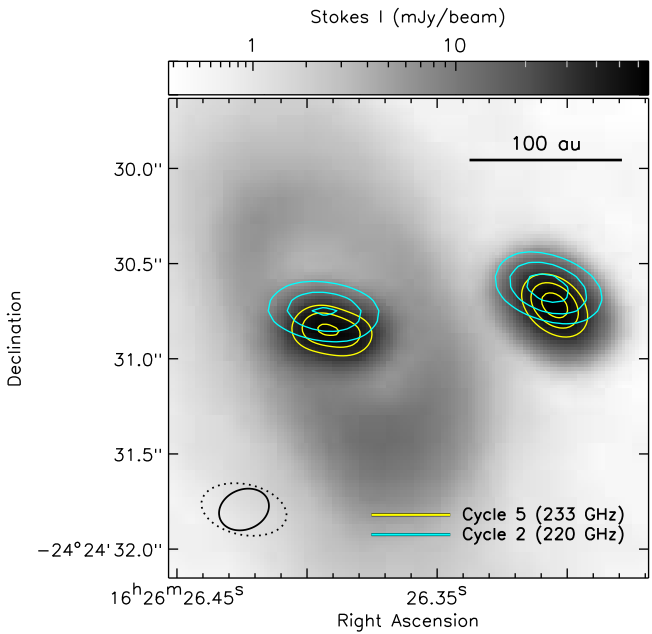}
\caption{Comparison of 1.3 mm continuum of VLA 1623 between Cycle 2 and the data presented here.  The background image shows our data. Contours correspond to 50 \%, 70 \%, and 90 \%\ of the peak intensity for our data (yellow) and the Cycle 0 data (cyan).  The resolutions of our data and the Cycle 2 data are shown in the bottom-left corner as solid and dotted ellipses,  respectively.  We use the label Cycle 5 for our data to represent the epoch of the observations.  The 220 GHz observations from Cycle 2 are taken directly from the ALMA archive.  \label{compare_cont}}
\end{figure}

VLA 1623 was first observed with ALMA in Cycle 0 at a resolution of $\sim 0.65$\arcsec\ \citep{Murillo13}.  It was also observed during Cycle 2 in multiple configurations, including the ACA.  The ACA data were published in \citet{Murillo15}.  For Figure \ref{dcop}, we use the highest resolution data from Cycle 2 (0.35\arcsec) that best matches the resolution of our polarization data.  Figure \ref{dcop} shows a slight positional offset between the 1.3 mm dust emission presented here and the \DCOp\ (3-2) absorption from the Cycle 2 data.  To verify this offset, Figure \ref{compare_cont} compares dust continuum from the Cycle 2 project (at 0.35\arcsec\ resolution) and our observations, which we label as ``Cycle 5'' to represent the observing epoch (recall that this is a Cycle 3 project).  The background image is a zoom-in of our 233 GHz continuum map of VLA 1623 with yellow contours and cyan contours outlining our data and the Cycle 2 data, respectively.   As with the \DCOp\ (3-2) absorption feature, we see a systematic shift of $\sim 0.14$\arcsec\ (half our beam) in the South-West direction.  

\begin{figure}[h!]
\includegraphics[width=0.475\textwidth,trim=1pt 1pt 1pt 1pt,clip=true]{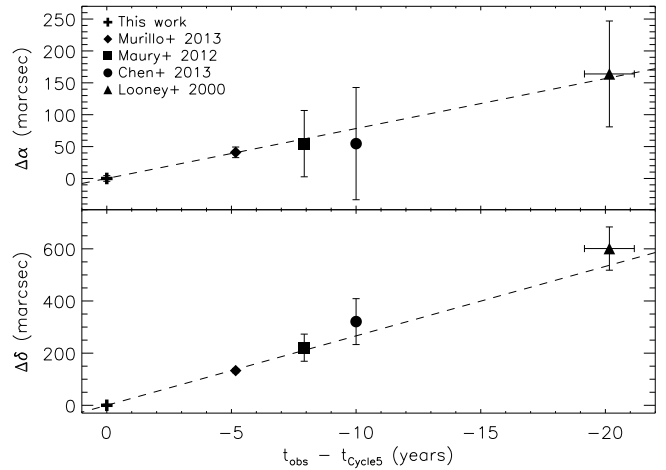}
\caption{Proper motion of VLA 1623-B in right ascension (top) and declination (bottom).  Position errors are estimated from scaling the geometric beam size by the peak intensity signal-to-noise ratio (see Appendix \ref{pm_app}).  For \citet{Looney00}, the epoch of observations is given only as a range from 1996 to 1998.  Dashed lines corresponds to $\chi^2$ linear regression fits.    \label{proper_motion}}
\end{figure}

Shifts of $\sim 0.1$\arcsec\ can arise from a number of factors (e.g., variations in weather at the time of observation, proper motion).  To compare our measured position of VLA 1623 with other measurements, we collect the positions of VLA 1623-B from four sub-arcscecond millimeter observations in the literature from ALMA \citep{Murillo13}, the Submillimeter Array \citep[SMA;][]{Maury12, Chen13} and the Berkeley-Illinois-Maryland Association (BIMA) millimeter array \citep{Looney00}.   Figure \ref{proper_motion} shows the relative position of VLA 1623-B between our data and the archival observations (see Appendix \ref{pm_app} for more details).  We use VLA 1623-B only because the archival data have lower resolutions and as a result, their positions of VLA 1623-A may be confused by the dust emission from the ring.  

Figure \ref{proper_motion} shows a clear trend in the position of VLA 1623-B in both right ascension and declination.  It is systematically moving south-west with time.  We attribute this change in position to the proper motion of VLA 1623 as a whole.  We fit the positions of VLA 1623-B with a $\chi^2$ linear regression (see the dashed lines in Figure \ref{proper_motion}) and obtain the following measurements for its proper motion, $\mu$,
\begin{eqnarray}
\mu_\alpha\cos\delta &=& (-7.8 \pm 1.6)\ \mbox{mas}\ \mbox{yr}^{-1}\ \mbox{and} \\
\mu_\delta &=& (-27 \pm 1.6)\ \mbox{mas}\ \mbox{yr}^{-1}.
\end{eqnarray}

\citet{Loinard08} measured the proper motion of two stars in Ophiuchus (S1 and DoAr 21) using VLBI observations.  Since the stars are also close to VLA 1623 on the plane of the sky (S1 is 2\arcmin\ away and DoAr 21 is 5.4\arcmin),  they could trace the same moving group.  Indeed, \citet{Loinard08} found that both stars are moving in the south-west direction, just like VLA 1623-B.  The proper motion of VLA 1623-B also has a similar magnitude as that of S1 ($\mu_{\alpha,S1}\cos\delta = -3.88$ mas yr$^{-1}$ and $\mu_{\delta, S1} = -31.55$ mas yr$^{-1}$), suggesting that our values fit with the overall movement seen in Ophiuchus.  

Thus, with a baseline of only three years with ALMA, we find a noticeable, systematic shift in the source positions of both VLA 1623-A and VLA 1623-B (see Figure \ref{compare_cont}).  As ALMA takes more data, embedded sources in other nearby clouds will likely show significant proper motions.  These proper motions will need to be accounted for in analyses that attempt to combine observations over different epochs.


\section{Summary and Conclusions}\label{summary}

We present new polarization data for VLA 1623 with unprecedented sensitivity and angular resolution.  These data are centered on the tight Class 0 protobinary containing VLA 1623-A and VLA 1623-B.  Our main results are:

\begin{enumerate}
\item We resolve compact, inner disk candidates with diameters of $\sim 40$ au around both VLA 1623-A and VLA 1623-B.  These observations are the first to resolve a disk candidate around VLA 1623-B.  We also resolve extended dust emission around VLA 1623-A as a ring.   This ring was identified as a disk in previous observations at lower resolution and matches the larger Keplerian disk identified from previous observations.
\item We see extensive and structured polarized emission compared to previous single-dish and interferometric observations.  Moreover, we find distinct polarization patterns toward the two inner, protostellar disk candidates and the ring.  The disks shown uniform polarization morphologies aligned with their short axes, whereas the ring shows azimuthal polarization.
\item We find that the polarization toward the inner protostellar disks is most consistent with dust scattering.  This result agrees with expected timescales of dust grain growth on $\lesssim 30$ au scales.  
\item The polarization for the ring is inconsistent with pure dust scattering.  We propose that the polarization may arise instead from grain alignment due to a magnetic field, an anisotropic radiation field, or a combination of the two.  
\item We find that the inferred magnetic field morphology is mostly radial with a hint of curvature.  This polarization structure does not match expectations for a toroidal magnetic field even though VLA 1623-A appears to have a large Keplerian disk.  We fit the radial magnetic field toward the ring with a simple, analytical model of a flux-frozen centrally-concentrated oblate spheroid.  Although this model does not include rotation, we find good agreement between the orientation of the model field and the observations.  Based on our model, we propose that VLA 1623-A has evidence of magnetization.   
\item If the polarization in the ring is due to magnetically-aligned dust grains align, then our model indicates that the magnetic field around VLA 1623-A has an hourglass morphology roughly parallel with the outflow axis.  More detailed models that include rotation, the parent core structure, and non-ideal MHD are necessary to fully address the magnetic field structure of VLA 1623-A. 
\item If the polarization pattern in the ring is instead due to grain alignment from the anisotropic radiation field, then VLA 1623-A may have experienced rapid grain growth to $\sim 80$ \um\ sizes at radial extents between $50-90$ au.  Comparisons with grain growth models over these radial extents are necessary to determine the timescales for producing $\sim 80$ \um\ grains over the Class 0 lifetime.
\item We find a half-beam positional offset between our observations and previous Cycle 2 ALMA observations of VLA 1623. We determine that this offset is from the proper motion of VLA 1623, which is moving in a south-west direction.  Corrections for proper motion may be important  when combining high resolution ALMA observations from different epochs for nearby molecular clouds.
\end{enumerate}

The dust polarizations observations of VLA 1623 presented here represent just the first results from a larger ALMA study of embedded sources in Ophiuchus.  Our observations showcase the ability of ALMA to detect dust polarization with high angular resolution and sensitivity in a very short amount of on-source time.  This ability makes  large surveys of dust polarization toward embedded stars possible.   These large surveys are important as polarization observations may be a useful diagnostic to disentangle compact emission from protostellar disks, outflow knots, and dense inner envelopes.

\vspace{1cm}
\begin{acknowledgements}
The authors thank the NAASC and EU-ARC for support with the ALMA observations and data processing.  The authors also thank Mark Reid for insightful discussions regarding interferometric pointing accuracy, Attila Kovacs for guidance in determining the deconvolved disk properties, and Sebastien Fromang and Patrick Hennebelle for theoretical discussions on magnetic fields in young systems.   SIS acknowledges the support for this work provided by NASA through Hubble Fellowship grant HST-HF2-51381.001-A awarded by the Space Telescope Science Institute, which is operated by the Association of Universities for Research in Astronomy, Inc., for NASA, under contract NAS 5-26555.  Research of Th.H. on moving groups and binaries is supported by DFG grant SFB 881 "The Milky Way System".  W. K. was supported by Basic Science Research Program through the National Research Foundation of Korea (NRF-2016R1C1B2013642). This paper makes use of the following ALMA data: ADS/JAO.ALMA\#2015.1.01112.S and ADS/JAO.ALMA\#2013.1.01004. ALMA is a partnership of ESO (representing its member states), NSF (USA) and NINS (Japan), together with NRC (Canada), MOST and ASIAA (Taiwan), and KASI (Republic of Korea), in cooperation with the Republic of Chile. The Joint ALMA Observatory is operated by ESO, AUI/NRAO and NAOJ.  The National Radio Astronomy Observatory is a facility of the National Science Foundation operated under cooperative agreement by Associated Universities, Inc.

\end{acknowledgements}

\bibliographystyle{apj}
\bibliography{references}

\begin{appendix}

\section{Proper Motion of VLA 1623} \label{pm_app}

VLA 1623-B has been observed at sub-arcsecond resolution by several studies in the literature.  For simplicity, we focus on only the (sub)millimeter observations, as radio emission can arise from jets and therefore, may not be centered on the protostar.  We find four studies in the literature that meet these criteria: \citet[ALMA, 1.3 mm;][]{Murillo13}, \citet[SMA, 1.3 mm;][]{Maury12}, \citet[SMA, 1.3 mm;][]{Chen13}, \citet[BIMA, 2.7 mm;][]{Looney00}.  These observations span 20 years.  

Table \ref{vla1623b} lists the positions of VLA 1623-B from our work and the literature.  The first column gives a shortened name for the reference (see above).  The second column gives the epoch of observations.  The third, fourth, fifth, and sixth columns give the coordinates and our adopted uncertainty for right ascension and declination, respectively. The seventh column gives the resolution.  All studies have elongated beams, so we use the geometric mean.  The eighth column gives the peak millimeter intensity and error.  The last two columns give the offset in right ascension and declination relative to our position. 

{\setlength{\extrarowheight}{0.8pt}%
\begin{table*}
\caption{Literature Positions of VLA 1623-B}\label{vla1623b}
\begin{tabular}{lclclccccc}
\hline\hline
Reference	&	Epoch	& $\alpha$ (J2000)	& $\sigma_{\alpha}$	 &$\delta$ (J2000) & $\sigma_{\delta}$	& FWHM\tablenotemark{a}	 & $I_{peak}$ & $\Delta\alpha$ & $\Delta\delta$ \\
	&			& (h:m:s)		&  (mas)			& (h:m:s)		& (mas)		& (arcsec)		& (\mJybeam)	& (mas)		  & (mas)	\\
\hline
This study	 & 2017.5		& 16:26:26.306	 &	4.5		& -24:24:30.721&	4.5		& 0.24		& 67.0 $\pm$ 1.3 & 0	  & 0		 \\
Murillo13\tablenotemark{b}	& 2012.33		& 16:26:26.309	 &	7.0		& -24:24:30.588&	7.0		& 0.65		& 93.5 $\pm$ 1 & 41	   & 133		 \\
Maury12\tablenotemark{c}	& 2009.58		& 16:26:26.31	 &	52		& -24:24:30.5	&	52		& 0.53		& 71 $\pm$ 7  & 55	   & 221	 \\
Chen13\tablenotemark{d}	& 2007.5		& 16:26:26.31	 &	88		& -24:24:30.4	&	88		& 0.42		& 71 $\pm$ 15 & 55	   & 321		 \\
Looney00\tablenotemark{e}	& 1997.34		& 16:26:26.318	 &	83		& -24:24:30.12	&	83		& 0.61		& 25.8 $\pm$ 3.5 & 164	   & 600		 \\
\hline
\end{tabular}
\begin{tablenotes}[normal,flushleft]
\item \tablenotemark{a} The geometric mean FWHM ($= \sqrt{ab}$).
\item \tablenotemark{c} \citet{Murillo13} does not quote an uncertainty for their peak flux of VLA 1623-B.  We use their map sensitivity as an estimate.
\item \tablenotemark{c} \citet{Maury12} quotes a position uncertainty of 30 mas.  We use our calculated value of 52 mas to be consistent.
\item \tablenotemark{d} \citet{Chen13} does not give the peak flux of VLA 1623-B.  We use the total flux from this study instead.  
\item \tablenotemark{e} \citet{Looney00} does not give the exact epoch of their observations of VLA 1623.  We use the average epoch from their entire survey for our analysis.  
\end{tablenotes}
\end{table*}
}

The uncertainties in right ascension ($\sigma_\alpha$) and declination ($\sigma_\delta$) in Table \ref{vla1623b} represent the estimated pointing uncertainty from each set of observations.  For simplicity, we assume $\sigma_{\alpha} = \sigma_{\delta} = \mbox{FWHM}/(I/\sigma_I)$.  This relation gives values ranging from roughly 4 mas for this work and \citet{Murillo13} to  roughly 90 mas for \citet{Chen13} and \citet{Looney00}.  These adopted pointing errors are likely overestimates, however.   \citet{Maury12} estimated a pointing accuracy of 30 mas using observations of their calibrators.  Similarly, we find typical accuracies of $\lesssim 1$ mas for our calibrators and $\lesssim 2$ mas between the three epoch observations of VLA 1623 in our data (see Section \ref{data}).  Nevertheless, we use the position errors from $\mbox{FWHM}/(I/\sigma_I)$ to be consistent with all the observations. 

We also consider two alternative causes for the positional offsets: the orbital motions of the stars and atmospheric noise.  We find that orbital motions are an unlikely cause.  First, the offsets are systematic (roughly equal for both VLA 1623-A and VLA 1623-B across the 20 year baseline), whereas orbital motions should produce opposite shifts for the two stars.  Second, VLA 1623-A and VLA 1623-B have projected separation of $\approx$ 150 AU.  Assuming 0.5 \Msun\ stars, the orbital period would be $\sim$ 1000 years.  Our observation baseline of 5-20 years is insignificant compared to a 1000 year period.  Therefore, we do not expect the binary orbit of the stars to have an effect on our measurements.  

Atmospheric noise, however, can produce systematic position errors of $10-100$ mas.  These errors are mitigated by long observations or observations taken over multiple days in different conditions so that the atmospheric fluctuations effectively cancel out.  Using the three epochs of our Cycle 5 observations, we found relatively minor positional errors of $\lesssim 2$ mas.  We do not have multiple epoch observations for most of the literature data, however.  Nevertheless, atmospheric variations will produce random offsets between different sets of observations, whereas we see systematic offsets over a 20 year baseline (see Figure \ref{proper_motion}).  Thus, we do not believe the offsets between our data and those in the literature are due to atmospheric variations.

\end{appendix}

\end{document}